\documentclass[5p]{elsarticle}

\usepackage{lineno,hyperref}
\modulolinenumbers[5]
\journal{Astronomy \& Computing}

\biboptions{authoryear}

\begin{document}

\begin{frontmatter}

\title{Understanding the human in the design of cyber-human discovery systems for data-driven astronomy}

\author[1,2]{Christopher J. Fluke\corref{cor1}}
\ead{cfluke@swin.edu.au}
\author[2]{Sarah E. Hegarty}
\author[3]{Clare O.-M. MacMahon}
\cortext[cor1]{Corresponding author}

\address[1]{Advanced Visualisation Laboratory, Digital Research  Innovation Capability Platform, \\
Swinburne University of Technology, John St, 3122, Australia}
\address[2]{Centre for Astrophysics \& Supercomputing, \\
Swinburne University of Technology, John St, 3122, Australia}
\address[3]{Sport and Exercise Science, School of Allied Health, Human Services, and Sport, La Trobe University, 3086, Australia}

\begin{abstract}
High-quality, usable, and effective software is essential for supporting astronomers in the discovery-focused tasks of data analysis and visualisation.  As the volume, and perhaps more crucially, the velocity of astronomical data grows, the role of the astronomer is changing. There is now an increased reliance on automated and autonomous discovery and decision-making workflows rather than visual inspection.
We assert the need for an improved understanding of how astronomers (humans) currently make visual discoveries from data. This insight is a critical element for the future design, development and effective use of cyber-human discovery systems, where astronomers work in close collaboration with automated systems to gain understanding from continuous, real-time data streams.   We discuss how relevant human performance data could be gathered, specifically targeting the domains of expertise and skill at visual discovery, and the identification and management of cognitive factors.  By looking to other disciplines where human performance is assessed and measured, we propose four early-stage applications that would: (1) allow astronomers to evaluate, and potentially improve, their own visual discovery skills; (2) support just-in-time coaching; (3) enable talent identification; and (4) result in user interfaces that automatically respond to skill level and cognitive state.  Throughout, we advocate for the importance of user studies and the incorporation of participatory design and co-design practices into the planning, implementation and evaluation of alternative user interfaces and visual discovery environments.  
\end{abstract}




\end{frontmatter}

\section{Introduction}
\begin{figure*}
\includegraphics[width=2\columnwidth,angle=0]{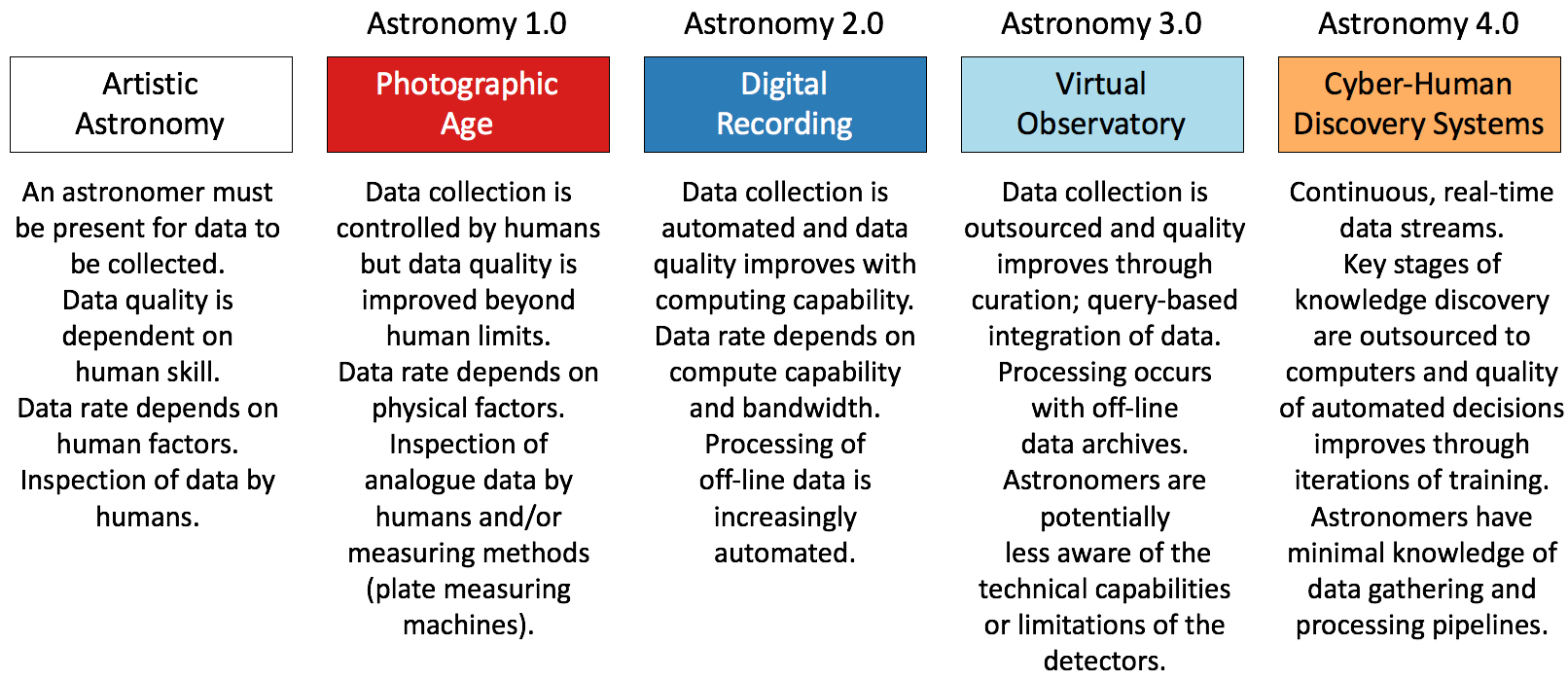}

\caption{The impact of technology transformations on the way astronomers work and make discoveries. Starting from the artist, working alone or within a small team, each technology transformation brings a change in the approach, the scale of the endeavour, and the role of the individual. }
\label{fig:astronomy4.0}
\end{figure*}

Progress in astronomy has relied on a series of inter-related processes: planning an observation or simulation; collecting or otherwise generating the relevant data; performing data analysis,  visualisation, model-fitting and related activities; and publishing and presenting results.  All of these processes are essential for the production and dissemination of discoveries, new knowledge, or insights. Individual processes are generally iterative, and do not necessarily occur in the linear order presented here.

Over time, there have been profound and fundamental changes regarding the role of the astronomer in conducting these processes -- particularly, but not exclusively, as they apply to observational astronomy.  The revolutionary impact of the telescope on astronomy notwithstanding, the way that data is collected and made available for analysis has been transformed multiple times (see Figure \ref{fig:astronomy4.0}).  Each stage has a resulted in an increase in the quantity and quality of data, and a reduction in the direct connection between the astronomer and the instrument \citep{Norris09}.

{\em Artistic} astronomy reached its peak with the hand-drawn sketches of Sir John Herschel, William Parsons (3rd Earl of Rosse), James Nasmyth, \'{E}tienne L\'{e}opold Trouvelot,  and their contemporaries \citep[see examples in the plates of][]{Ball1900}, coincident with the dawning of the {\em photographic age} in the 19th century \citep{Barnard1898a,Barnard1898b}.  Photographic plates and analogue chart-recorders entered obsolescence with the move to fully {\em digital recording} devices \citep[e.g.][and references therein for the history of charge-coupled devices in astronomy]{Lesser2015}.  

As digital data became easier to share, the use of dedicated data archives emerged [e.g. through the {\em Virtual Observatory} and related online archives \citep{Brunner98,Szalay99,Brunner01,Quinn04}], along with presenting opportunities to develop, adopt and apply data mining and automated discovery methods.

The growth in both the quantity ({\em volume}) and rate of data ({\em velocity}) from new astronomical instruments, sensors and numerical techniques presents both an increased discovery potential and a challenge to astronomers to ensure they make full use of their data \citep{Berriman2011}.

\subsection{Visualisation for discovery}
The human visual system is regularly proposed as being {\em the} gold standard\footnote{A concept emerging from medical diagnosis, a gold standard may not be an indicator of absolute ground truth, but is the best available method at the time \citep{Versi92,Claassen05}.}   for novel discovery in astronomy, but with limited research to support that claim. In other fields, more attention has been paid to the cognitive and visual effects that may limit human capabilities \citep[e.g.][for assessment of methods for automating digital tissue image analysis in pathology]{Aeffner17}, and cases where automated methods may out-perform human capabilities [e.g. \citet{Hooge18}, and references therein, and \citet{Hosny18}, for an overview of advances in artificial intelligence in radiology].

Many observing programs are predicated on their potential to uncover ``unknown unknowns'' by opening up previously unexplored regions of parameter space \citep{Harwit03}. Indeed, as \citet{Norris17} states, {\em ``most major discoveries in astronomy are unplanned''}, indicative of the  advantages of approaching data with an open mind. However, future success in discovering the unexpected actually requires a great deal of pre-planning today, and many strategies require knowledge of ``known knowns'' as templates.  See also \citet{Fabian09} on the need to be sufficiently prepared for serendipitous discovery in astronomy. 

In many instances, discovery of the first exemplar of a new astronomical object or new phenomena \citep{Norris17} has relied on visualisation, for example, looking at a photographic plate or CCD image, the output of a chart recorder \citep[e.g.][]{Hewish68}, or identifying complex structure in a scatter plot, such as the ``stickman'' in the Center for Astrophysics redshift survey \citep{deLapparent86}.

As data continues to be created more rapidly, there is a corresponding reduction in the time available for visual inspection of individual spectra, two-dimensional images, three-dimensional data cubes, and a host of other derived and related multi-dimensional data products.    In the past, it was feasible that a majority of photographic plates, digital pixels or voxels recorded {\em could} be inspected by a human. Annie Jump Cannon, Williamina Fleming, Henrietta Leavitt, and other `human computers' diligently reviewed around half a million plates while employed by the Harvard College Observatory \citep[e.g.][]{Nelson08}.  Through online citizen science projects, most notably Galaxy Zoo\citep{Lintott08}, visual analysis and classification has been successfully outsourced, increasing the number of human inspectors well beyond the membership of the research teams who gathered the data \citep[see also the review by][]{Marshall15}. Looking ahead, this opportunity to view everything is almost entirely eliminated.  Indeed, the expectation is that the vast majority of new astronomical data, from observation and simulation, will {\em never} be looked at by a human.\footnote{If we consider the Square Kilometer Array (SKA) as the benchmark, generating 300 PB/yr \citep{Scaife20}, and conservatively estimate 2--3 PB/yr would be directly inspected, then $> 99 \%$ of this data volume will never be viewed by astronomers.}

This is a new stage in the evolution of astronomy and astronomers.   Here, continuous processing of tera-, peta-, and exabyte scale data streams will be required, where intelligent autonomous systems learn to identify salient features in data, and present them to astronomers for confirmation and customised analysis.  
Human capabilities for knowledge discovery and insight will be augmented through an ever tighter integration with intelligent computing systems.   Through the use of cyber-human discovery systems\footnote{This name deliberately echoes the emergence of cyber-physical systems in Industry 4.0, where machines interoperate with physical processes, making real-time, data-driven decisions.}   (see right-hand column of Figure \ref{fig:astronomy4.0}), the next generation of astronomers will work in ever-closer partnership with the ``machines'' to make discoveries and advance knowledge of the Universe. 

\subsection{Cyber-human discovery systems}
While increased reliance on automated discovery seems inevitable, in the near-term, human intervention will continue at all stages of the data collection and processing cycle.  Visual inspection tasks will include quality control (``is the data usable?''); validation (``did the source-finder work effectively?''); and discovery (``are these objects consistent with previous discoveries or do they represent a new type of object?'').     At the same time, reducing the reliance on slow, error-prone, inefficient human cognitive capabilities by handing ever more complex discovery and decision-making tasks to machines, may have a positive impact on scientific progress overall \citep{Gil14}.  

Our ability to visualise data, draw insight, and make decisions is limited by the capabilities of the technology we use to complete these tasks.  High-quality, usable, and effective software is essential for supporting astronomers in the discovery-focused tasks of data analysis, visualisation and visual analytics.  Additionally, a variety of different data displays and interaction devices are available, yet most astronomers still work mostly with a desktop or laptop-style monitor.  

\citet{Rots86}, \citet{Norris94}, \citet{Richmond94}, \citet{Gooch95}, and \citet{Fluke06} have all been optimistic about the availability of technology and its potential to improve the way that we visualise and interact with data in astronomy.   Using a solution because it is available (e.g. through habit or based on training that was available) does not necessarily mean it is the most appropriate option, or provides the greatest opportunity for an individual to work at their maximal skill level.   Contrast the different visual experiences provided by a low-resolution smart-phone screen held at arm's length with that of a 4K projection in a dedicated collaborative visualisation workspace.\footnote{Smart phone users need not despair, as individuals can and do learn to work effectively with the limitations of a device, and continue to make discoveries.  Understanding of this {\em facsimile accomodation} emerged in studies comparing viewer responses to original large-scale artworks and small-scale reproductions \citep{Locher99}.} Additionally, the nature of a display may contribute, positively or negatively, to cognition, understanding and training: people learn better in conditions that have greater contextual interference and require more engagement \citep[e.g.][]{Magill90,Barreiros07}.

There is a need, therefore, to ensure that the astronomer's discovery potential is maximised for processes where the human visual system is still actively engaged.  At the same time, we need to recognise that there are important differences in the way individual astronomers see and interpret data \citep[e.g.][]{Schneps2007,Schneps2011} and make decisions about, or discoveries, from data.  Additionally, such a strong focus on visual discovery does not recognise the potential to use multi-sensory methods, such as data sonification, which can better engage vision-limited scientists [e.g. \citet{Candey05} and see Section \ref{sct:sonify}]. This topic of the personalisation of data-driven discovery environments has been subject to only limited systematic investigation by the community that it affects the most.  
 
 By understanding more completely how astronomers make discoveries -- or want to make discoveries -- and linking this to factors such as experience level, skill, or cognitive state, improvements could already be made to individualise the display of data.  This could include user experiences that accommodate natural interaction experiences (gaze-based interaction, hand gestures rather than mouse-based navigation) or promote easier and more wide-spread use of emerging display technologies (virtual, augmented and mixed reality as prime examples), perhaps by simplifying the data interchange processes \citep{Fluke09,Vogt16}.

The majority of astronomy visualisation software does not specifically provide a method to optimise  -- automatically or manually  -- the display of data to an individual's strengths, skill level, cognitive load or physical state.   Future user interfaces for astronomers could  measure and adapt to the cognitive or physical state of the astronomer, providing additional content to an individual in anticipation that the user was entering a phase of high discovery potential, or limiting the data flow as the astronomer becomes cognitively overloaded and needs a break.

\subsection{Overview}
In this paper, we consider two aspects of the human component within future cyber-human discovery systems for astronomy:
\begin{itemize}
\item {\em Expertise and skill}: the ability to complete a visual-discovery or decision-making task depends on the astronomer's expertise (most easily linked to career stage or familiarity with a task) and skill level (dependent on inherent `talent' at performing visual processing tasks); and
\item {\em Cognitive and physical factors}:  regardless of expertise or skill level, other factors can impose time-dependent variations in proficiency, such as attention (the astronomer is focused on all aspects of the task), workload (the astronomer is in control of when they are completing tasks), and cognitive state (the astronomer has sufficient cognitive capacity and is not fatigued).
\end{itemize}

We propose the need to gather human performance data in order to establish a baseline of skill level and cognitive factors in astronomers when performing visual discovery tasks.  Such data is relevant for an improved understanding of how the needs of astronomers will be supported within cyber-human discovery systems. We explain how this could be achieved with off-the-shelf hardware and software.

By understanding skill level and cognitive factors, improved user interfaces could be developed that respond automatically to the individual. This could be achieved through provision of just-in-time coaching or adaptations of the way visual content is presented to maximise the potential for an individual to make a discovery in a data-intensive/data-streaming context.

Additionally, we advocate for increased use of participatory design and co-design practices, to improve the user experience and suitability of visual displays of data in all of their forms.  

\section{Background and related work}
\label{sct:background}
In this section we provide an overview of related research linked most closely to studying human factors in astronomy, and the utilisation of participatory design methods for developing data visualisation interfaces.  We introduce and define the complementary dimensions of expertise and skill as they relate to performing visual discovery tasks.

\subsection{Augmented intelligence and human-in-the-loop}
\citet{Zheng17}, amongst others, writes about the emerging close relationship between the human and the machine in terms of (hybrid-){\em augmented intelligence}.   The augmentation can occur through a cognitive model, where understanding of the human brain, including biological factors, is used to develop new software and hardware which more closely replicates the way a human would think about or solve a problem.   

Alternatively, the augmentation requires a human-in-the-loop.  Here, human judgment and critical reasoning is applied to the outputs of an intelligent system, e.g. through training and validations steps, thus raising the level of confidence in the outcomes of the artificial intelligence.   For related work, see, for example, \citet{Borne09} for a discussion of  human and machine collaboration for annotation of features in data streams, and \citet{Fuchs09}, examining how interactive visual analysis can be used to help steer a machine learning system to reach a more reasonable hypothesis. 

 \citet{Rogowitz12} proposed a framework for human-in-the-loop discovery. They highlighted the continuous and close coupling between the user's decisions and judgements, and the algorithmic processes that transform and modify the data.  The user steers the discovery process by selecting regions of interest, which can be represented with a variety of visualisation techniques most appropriate for the data.  Once a feature of interest is identified, an algorithmic step seeks to obtain a mathematical description, paving the way for a more targeted search for similar features in a larger dataset.   This approach requires the active presence of the user in the initial knowledge discovery stage, however, the user must still have the cognitive capacity and relevant experience to identify interesting and important features.   
 
\subsection{Human factors in astronomy}
In order to design effective cyber-human discovery systems, we need to understand more about the human aspects of astronomy.   Human factors research is well established in other fields, ranging from talent identification and coaching for elite sport performance; expert diagnostic analysis and inspection of medical imaging; aviation, including attainment of flying skills and air traffic control operations; fire command and control; and military operations. While individual astronomers develop an intuition about how discoveries are made in their own field, there have been few investigations into human performance factors in astronomy more generally.

Social scientists \citep{Garfinkel981} analysed the process of discovery using a tape recording from 16 January 1969.  On that night, John Cocke and Michael Disney conducted observations at Steward Observatory, culminating in the first optical detection of a pulsar \citep{Cocke1969}.  The recording enables a comparison between the actions taking place at the telescope, with the documentation of the discovery in the resultant Nature publication.  While the technical description of the discovery in \citep{Cocke1969} is presented in the voice of the `transcendental analyst' adopted by most scientific authors, \citep{Garfinkel981} witnesses the `first time through' nature of the observing run as the `shop practices' of two astronomers result in a potential discovery turning into a reality.   The discussion between the astronomers is as much about the importance of the discovery as it is about the properties of the pulsar itself.   

A pioneering effort to provide a cyber-human discovery system in astronomy was completed by \citet{Aragon08a,Aragon08b}.  Their Sunfall Data Taking system  supported a team of astronomers to make collaborative, real-time decisions when observing for the Nearby Supernova Factory (SNfactory) project \citep{Aldering02}.  The SNfactory collaboration utilised the SuperNova Integral Field Spectrograph (SNIFS) on the University of Hawaii 2.2m telescope (Mauna Kea, Hawaii), producing a nightly data rate of 50-80 GB which needed to be processed within 12-24 hours.   

\citet{Aragon08a} compared the role of an observational astronomer to that of a pilot: working at altitude, at night, responding to a variety of rapidly changing weather conditions while ensuring safety of staff and a multi-million dollar asset.  In aviation and the aerospace industry, this ability to make time-pressured decisions that require detailed attention and an understanding of cause and effect, is referred to as {\em situation awareness}.   Endsley's widely-used model of situation awareness \citep{Endsley95} comprises three stages: (1) perception of the current state of the important factors in the environment; (2) comprehension and synthesis of information to gain a clear overview; and (3) projection or prediction of what will happen next within the environment, based on knowledge of likely patterns.

Using the framework of situation awareness to inform the design, the Sunfall Data Taking system helped reduce human errors, such as failing to follow-up targets.  Leveraging principles from participatory design (see below), Sunfall combined visualisation,  machine learning and data management, paying careful attention to the deployment of a graphical user interface that minimised cognitive load.  User evaluation, through interviews and analysis of logs, showed an overall positive impact of Sunfall on improving situation awareness, efficiency and collaboration.

 Concerned by the lack of evidence that tiled display walls (TDW\footnote{A TDW is constructed by combining smaller, commodity monitors to create an ultra-high-resolution display.}) had a genuine role to play in astronomy, \citet{Meade14} recruited 45 non-astronomers and 12 astronomers to participate in a series of image search tasks.     Individual and collaborative inspection was investigated, with tasks performed on both a standard desktop display ($1680 \times 1050$ pixels) and the OzIPortal at the University of Melbourne (comprising 24 flatscreen LCD monitors, each with $2560 \times 1600$ pixels, arranged in a $6 \times 4$ matrix for a total of $15360 \times 6400$ pixels). Small features were identified in images on the standard display by panning and zooming, while physical navigation was used to walk around and view different regions on the TDW.  Results were reported in terms of search success rates (i.e. how often a target was found) and analysis of a post-test survey.    
 
 While the search success rates for both the standard display and TDW were comparable, with astronomers outperforming non-astronomers, the post-test survey revealed that both cohorts felt that the TDW was easier to use for the image search task and was more suitable.  These insights provided the impetus to use a TDW as part of a wider display ecology for collaborative, real-time data exploration for the Deeper Wider Faster rapid transient search project \citep{Meade17,Andreoni19,Andreoni20}.

\subsection{Eye tracking}
\label{sct:eyetrack}
Often used as a method for identifying individual differences in the inspection of images, eye tracking methods \citep{Yarbus67,Duchowski07,Tatler10,Holmqvist12} have been used sparingly in astronomy.    This may be due to a lack of awareness in the astronomy community, an inability to access and experiment with eye tracking solutions, or simply be a missed opportunity for interdisciplinary collaboration.

By recording where, when and for how long viewers look at different parts of an image, insight can be gained on how a particular visual display of information is interpreted, and whether an alternative representation of data might be more effective.  In many visual processing tasks (c.f. anticipation or ``reading the play'' in sport), eye tracking permits a move to {\em process} measures and not just outcome measures (e.g. accuracy, completeness), i.e. how the visual activity was performed, not just what the outcome was.

Two main methods for presenting and interpreting eye-tracking data are {\em attention maps}, which measures the accumulated time spent looking at different parts of an image, and {\em gaze plots}, which provide spatio-temporal information of how a viewer's gaze moves around an image.  See \citet{Kurzhals16}, and references therein, for a comprehensive review of eye tracking research within the field of visual analytics.

Based in part on eye-tracking experiments, \citet{Schneps2011} found evidence that individuals with dyslexia may have a neurological benefit when it comes to identifying features in image-based data.  While dyslexia is usually associated with difficulties with reading, other aspects of visual processing are potentially enhanced -- such as an ability to identify symmetric signals in noisy data, which requires a higher level of peripheral to central visual processing \citep[see also][]{Schneps2007}.

\citet{Arita11} completed a user study with 20 participants, to examine whether there were differences in the gaze patterns between four expert astrophysicists and sixteen novices (i.e. with limited background knowledge of astronomy).  The goal was to determine whether novices and experts looked at images differently, and whether this could be used to create visualisations that were better able to draw the attention of one or both cohorts. 

Participants were shown a sequence of Hubble Space Telescope images, along with simulation images created with the Spiegel visualisation framework \citep{Bischof06}, and their gaze patterns were recorded with a Mirametrix S1 eye-tracker.  Tasked with providing a verbal description of the quality of each image, the eye-tracking data was presented as an average fixation duration.  The expert cohort spent slightly more time focusing on a smaller region of each image, whereas novices tended to scan a larger part, but the two results were highly correlated.  In the \citet{Arita11} study, the open-ended nature of the visual task -- based on a self-assessment of quality -- meant that specific visual search and discovery strategies, either within or between the groups, were not examined.

\subsection{Participatory design}
By necessity, most visualisation software and algorithms for use in astronomy have also been implemented by domain experts in astronomy.  Consequently, developers often have limited understanding of best practice in supporting human computer interaction.  This can lead to applications for data exploration, visualisation and discovery that are rich with features, highly-customised to perform the data analysis tasks required by astronomers, but which are low on usability or accessibility by particular user cohorts.

Participatory design (or co-design) is a truly collaborative process between designers, developers and domain experts.  The aim is to understand more fully what the user does or wants to be able to do, in order to create a solution that more closely meets these goals than occurs in more traditional design (requirements are gathered and a solution is delivered).  Participatory design often involves a cycle of design iterations, where prototypes are constructed, used, evaluated, and improved.

Early discussion of the design of graphical user interfaces for astronomy include: descriptions of fundamental principles \citep{Pasian91}; commentary on the state of data analysis systems in astronomy and the need for new solutions to cope with the equally-relevant data volume, velocity and variety challenges of the time \citep{Pasian93}; and the potential for  portable or multi-platform interfaces, as presented through the StarTrax-NGB interface for the High Energy Astrophysics Science Archive \citep[HEARSARC;][]{White93,Richmond94b}.

In astronomy, examples of participatory design such as in \citet{Aragon08a} -- where there is an emphasis on usability of the system achieved through collaborative design between the end-users and the software developers -- are rare or rarely documented.  \citet{Bertini93} and \citet{Pinkney94} refer to their use of participatory design to develop a Visual Browsing Tool (VBT) integrating with the Astrophysics Data System, that could be used to explore heterogeneous data collections via a visual query language.

\citet{Schwarz11}, \citet{Pietriga12}, and \citet{Schilling12}  made use of multiple participatory design workshops to design a user interface for the complex monitoring operations of the Atacama Large Millimeter Array \citep[ALMA;][]{Brown04} radio-telescope.  A solution that reduced the cognitive load of users, and supported clear and rapid decision making -- particularly regarding critical incidents -- was essential.    Visual and easily-accessible geographical data (i.e. locations of the moveable antennas), antenna status, and the resultant impact of a faulty antenna on baselines were all identified as improvements to the original control system design.   The methodology used in planning the ALMA operations control system was adopted for prototyping of a potential web-based, and hence remotely-accessible, user  interface for the Cherenkov Telescope Array \citep{Sadeh16}.

Assessing the suitability of existing three-dimensional visualisation tools for use in interactive analysis of radio astronomy spectral data cubes,  \citet{Punzo15} enlisted 15 participants in a review of four alternatives: {\tt Paraview}\footnote{https://www.paraview.org}, {\tt 3DSlicer}\footnote{https://www.slicer.org},  {\tt Mayavi2}\footnote{https://github.com/scibian/mayavi2},  and {\tt ImageVis3D}.\footnote{http://www.sci.utah.edu/software/imagevis3d.html}  Each user spent one hour working with the four options, and then provided feedback on usability, considering factors such as intuitiveness of the user interface and suitability for typical spectral cube analysis tasks. At the end of this user study, {\tt 3DSlicer} was selected as the most promising for further enhancement, resulting in the {\tt SlicerAstro}\footnote{https://github.com/Punzo/SlicerAstro/wiki} project \citep{Punzo16,Punzo17}.

\citet{Rampersad17} describes an iterative design process, starting on a foundation of user requirement gathering and paper-based prototyping, prior to implementation of a graphical user interface that could be evaluated on the basis of its usability.   The outcome was a proposed interface for visualisation of data cubes with an improved aesthetic quality (e.g. greater use of on-screen icons),  compared to other existing solutions.  This resulted in more intuitive ways to access functionality, which could potentially improve the experience for new learners by limiting the need for users to remember complex task sequences.  By engaging with domain experts at various stages of the design process, the prototyping process also identified new modes of interaction that would improve visualisation and analysis workflows, such as comparing data cubes or exporting high-fidelity images.

In most cases, the interface development occurs in stages, with input from experts and evaluation of the usefulness or success of the ideas that are generated.   A potential problem, though, is the lack of experts to take part in the participatory design or user studies \citep[e.g.][]{Meade14,Punzo15,Rampersad17}.  

A second pitfall is that not all experts will agree on the optimal design,  as all experts, and indeed, all astronomers are individuals. 
Writing in the context of access to data archives, \citet{Pasian93} identified the need for a collaborative approach between astronomers and computer scientists: ``{\em user interfaces need to be designed by astronomers, and possibly implemented by industry, but in close contact with the astronomers themselves...so as to follow the way of thinking which is culturally shared by the community of users.''}  However, a single {\em culturally shared} solution, by design, does not allow for variations in the presentation of data that would better suit or support individuals in their access to, exploration of, or analysis of data.

Finally, in order to identify relevant experts, we need to articulate the characteristics that suggest an astronomer has {\em expertise}, while recognising that there is another dimension relating to an individual's underlying, latent or natural {\em skill} level.

\section{Understanding skilled performance}
\label{sct:exskill}
Expertise can be gained simply by repeating a particular task a sufficient number of times, such that the required steps become automatic.   But there is more to being an expert than proficiency at a task.  Experts will generally possess a broader background knowledge, and can hold multiple mental models or competing hypotheses about the information presented to them before they make a decision. For a comprehensive study into the many facets of expertise, see \citet{ericsson18}.

For \citet{Arita11}, the threshold for expertise was low -- defined as having completed a Masters degree in astronomy. Such a definition of expert is not suitable for selecting a specific set of visual discovery skills that could best be supported or enhanced within an individualised or adaptive cyber-human discovery system.  A similar problem was faced in the \citet{Meade14} TDW user study, where the expert category comprised research astronomers, but not necessarily including individuals who possessed well-developed visual search strategies for images.    To understand what an expert does, we need to clearly define the type of expertise we aim to explore, and then gain the participation of astronomers who possess that expertise.

\begin{figure}[ht]
\includegraphics[width=1.00\linewidth,angle=0]{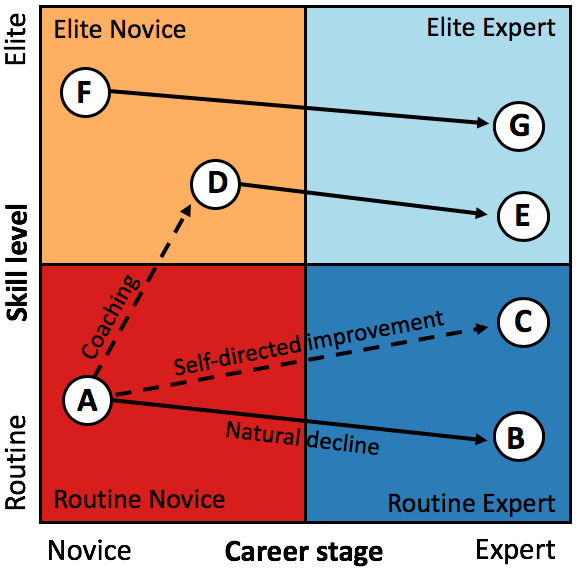}
\caption{Skill level and career stage.  We capture the two dimensions of expertise, which is linked to career stage, and skill, identifying four quadrants: routine novice, elite novice, routine expert, elite expert -- see Section \ref{sct:exskill} for details.  Potential career trajectories are discussed in Section \ref{sct:coaching}.  A novice may already possess an elite talent-level (F), or a higher skill level may be obtained without external intervention through self-directed improvement (C).   With the aid of coaching,   skill level may be boosted to a higher level, such that elite potential can be unlocked and utilised (D).   Over time, there may be some level of degradation in skill level, or the definition of what attainment of an elite talent level may change (career trajectories leading to B or G).   Without an objective test, or exposure to an opportunity to apply skills, an individual's latent skill level may not be identified until some way into her/his career (D).  }
\label{fig:ExpertCareer}
\end{figure}

Possessing expertise (i.e. knowing how to complete the task) does not necessarily ensure that an individual also possesses a high or elite skill level (i.e. an ability to complete a task measurably faster, more accurately or more effectively than another person -- including those who may have a comparable, or higher, level of expertise).   An individual with elite-level talent might be expected to identify features of interest  more readily than another astronomer with a lower skill level. However, an astronomer with latent elite skill at a particular visual discovery task might not be able to discriminate between known categories of sources, noise and spurious signals, or accurately identify ``unknown unknowns'' -- activities that rely more on expertise.

\citet{Hatano86} framed expertise in terms of {\em routine expertise}, e.g. knowing how to complete tasks competently according to a set procedure,  and {\em adaptive expertise}, where there is greater flexibility in thinking and problem-solving particularly when approaching new or novel scenarios \citep{Carbonell14}.   Without on-going training, or {\em deliberate practice} [see, for example,  the work of \citet{ericsson07,Ericsson08} and \citet{Ericsson10}, with a focus on expertise and superior performance in clinical medicine], the routine expert may never evolve to become an adaptive expert \citep{Chi2011,Carbonell14}.

Within the context of visual discovery by astronomers, the focus of this work, we capture the two dimensions of expertise and skill in Figure \ref{fig:ExpertCareer}, identifying four quadrants:
\begin{itemize}
\item Routine novice: a newcomer to visual discovery, highly competent, but not showing signs of elite-level capabilities;
\item Elite novice: a newcomer to visual discovery, with latent skills that allow them to operate at an elite level or unlock this elite potential through coaching; 
\item Routine expert: achieves expert status through repeated experience with visual discovery activities, but skill level never reaches the elite level; and
\item Elite expert: combines the attributes of expertise and elite skill, which likely include attributes of adaptive expertise.
\end{itemize}

Consider the case of sports talent identification.    Two athletes may have the same physical capabilities and fitness, as measured by vertical jump heights, performance on beep tests, or by measuring muscle fibre.   These results are usually indicators of elite potential, as they provide a quantitative way to separate the casual or active participant in a sport from a higher tier of achievement.   However, additional factors may set the athletes apart, based on psychological or cognitive attributes: which athlete makes better decisions more often under pressure?  In the long term, such factors might be more indicative that the athlete will demonstrate elite skill in a variety of conditions.

In astronomy, there are no equivalent measures of latent visual-discovery skill.    A common approach to training involves a graduate student being shown how to perform a visual discovery task by a supervisor or collaborator, and then left to achieve the task through trial and error.   This well-established method for research training allows for progression from novice to expert over a career, but may not be sensitive to, or unlock substantial growth in, skill level.     

A comparison of expert and novice perception and understanding of astronomical images was undertaken by \citet{Smith11} using a combination of online surveys and focus-group discussions.  Factors considered included the use of colour, the presence of text, and whether having background stars in images influenced the level of understanding.  Gathering data from an extensive cohort (8866 responses), it was found that text descriptions enhanced comprehension,  and that there were differences between the novices and experts (astrophysicists) in the aesthetic judgements regarding the use of colour.

\citet{Fluke2017} undertook a pilot study in performance analysis for visual discovery using the {\sc SportsCode}\footnote{https://www.hudl.com/elite/sportscode} software.  Originally developed by Sportstec, {\sc SportsCode} is a sports performance and coaching tool offered by Hudl.\footnote{http://www.hudl.com}   Presenting a slice-by-slice animation of a radio spectral data cube, an expert astronomer was able to apply descriptive text-based codes and graphical annotations to signal and noise components.   Moreover, they identified a need to better understand talent identification, e.g. for visual discovery in astronomy, as a way to provide more targeted training and coaching of astronomers in the data-intensive era.

\section{Gathering human performance data}
\label{sct:cyber}
A simple approach to gathering information on novice and expert behaviours and processes for visual discovery is to observe astronomers in action [see Table \ref{tbl:seven} and the summary in Section \ref{sct:seven} for methods of assessing user interactions, based on \citet{Lam12}]. Unfortunately, this is a labour intensive process, and likely restricted to a small number of participants or a very localised setting.  Instead, we look to technology-based approaches, that could be integrated into a visual discovery workflow -- particularly in cases where low cost, off-the-shelf solutions can be adopted. 

We consider four main options: (1) user interactions and log analysis; (2) think aloud and verbal reporting using speech-to-text conversion; (3) eye tracking; and (4) biometric sensing of cognitive and physical factors.

\subsection{User interactions and log analysis}
Automatic recording of user interactions with visual discovery interfaces provides insight on the sequence and duration of tasks.  In some instances, time-stamped log files are generated based on user operations with data, which could be inspected to determine how long particular tasks take to be completed by different user cohorts. However, it is rare for established astronomy visualisation software to include more detailed records of every event during an interactive session.

By deploying a visual discovery workflow in a web-browser, it becomes easier to automate the collection of additional user interactions.  This can include  time-based recording of mouse actions (movements, clicks, scrolling, etc.), text entry, and the sequence within a user interface that tasks are performed.   

This approach was used extensively in the Deeper Wider Faster (DWF) observing campaign.   DWF is a geographically-distributed, temporally-coordinated, multi-wavelength program searching for very short duration transient events in real-time \citep{Andreoni19,Andreoni20}.  During each campaign, a stream of ``postage stamp'' difference images of transient candidates is generated by the {\tt Mary} pipeline \citep{Andreoni17} and presented to a team of astronomers for real-time inspection and classification.   

Thus, a fundamental aspect of DWF data analysis is near continuous, real-time decision-making as to whether candidates are interesting (i.e. fast transients for which immediate follow-up observations are actioned) or not (e.g. asteroids, known variable stars, processing artifacts).   \citet{Meade17} provide a description of the display ecology used to support this visual inspection during early DWF campaigns.    

As DWF evolved, however, there was a need for a bespoke inspection and classification system that provided users with immediate access to difference images, light curves, and other transient diagnostics. Accordingly, the browser-based {\tt PerSieve} platform was developed (Hegarty et al. {\em in prep}).  {\tt PerSieve}  is integrated with the DWF processing pipeline via a PostgreSQL\footnote{\url{https://www.postgresql.org}} candidate management database, which is updated by {\tt Mary} in real time.  {\tt PerSieve}'s web portal then uses the Bokeh\footnote{\url{https://docs.bokeh.org}} graphics library to provide interactive in-browser visualisation and assessment of incoming transient candidates.  This approach allows continuous, user-based logging of interactions with the {\tt PerSieve} interface, which can give significant insight into the users's visual discovery workflow.

During two of the DWF observing campaigns (February and June 2018), particular attention was placed on studying user expertise, with each user self-rating as either  novice, intermediate or expert.  While a full description of the software, data collection and interpretation is to be presented elsewhere (Hegarty et al. {\em in prep}),  the {\tt PerSieve} experience showed that novice and expert users approached their decision-making in a very different way.  A preliminary investigation of time-based interaction workflows with the user interface showed that, in general, experts were able to definitively classify objects as interesting/not-interesting more quickly than novices. Missing from this study was sufficient information on why or how these decisions were made.

\subsection{Think aloud and verbal reporting}
An active way of gathering the ``whys'' and ``hows'' of visual discovery is to get astronomers to talk through their processes.  This can be performed while the task is being completed -- {\em think-aloud} protocols [\citet{Ericsson80,Ericsson98}, and compare with the specific case of novel discovery in astronomy as reported by \citet{Garfinkel981}] -- or as a summary at the end of the task -- a {\em verbal report} \citep{McPherson00}.

A challenge of any verbal description is the presence of a suitable vocabulary.   Consider the case of looking for ``unknown unknowns'' -- how do you get a novice or an expert to articulate what they are looking for when they may not have a word for it?  Moreover, a wider vocabulary, as might be expected of an expert, can lead to verbal overshadowing.\footnote{For example, some wine experts may just have more words, not more actual taste skills than others.}  For analysis of think-aloud or post-task verbal reports to be most useful as a means of distinguishing novice and expert behaviour, there is a need for astronomers to eschew obfuscation, while also being true to accepted usage of astronomical jargon and the individual's personal language idiosyncrasies.

As with the limitations of in situ observations of astronomers at work alluded to at the start of this Section, there is a need to automate the collection and transcription of audio.   This can be achieved using speech-to-text services, however, the quality of the transcription can depend strongly on the rate of speech, language, the audio quality of the environment, and the type of domain-specific vocabulary employed.   Closed-source and commercial solutions may perform more robustly than open-source alternatives, but the availability of cloud-based speech-to-text application programming interfaces (APIs), such as Google Speech-to-Text\footnote{\url{https://cloud.google.com/speech-to-text}}, Microsoft Cognitive Services\footnote{\url{https://azure.microsoft.com/en-us/services/cognitive-services/}}, IBM Watson\footnote{\url{https://www.ibm.com/cloud/watson-speech-to-text}}, offer a sufficient amount of flexibility.   

Automation of the interpretation of transcribed audio can also be supported, for example, through the methodological framework of computational grounded theory \citep{Nelson20}.  Here,  a multi-stage process of computer-aided content analysis, which can include machine learning and natural language processing, starts with the detection of important features in the text, followed by further refinement of the identified themes, and concludes with a pattern confirmation step.   

\subsection{Eye tracking}
Astronomical visual discovery is performed by having an astronomer look at data.   As introduced in Section \ref{sct:eyetrack}, eye tracking technologies can be used to determine where, when, and in what order different parts of an image are viewed -- and which parts are ignored.  

Assessing the wide variety of eye-tracking technologies that exist is beyond the scope of this article. In general terms, eye-tracking can be achieved using wearable devices integrated within glasses, as a peripheral camera, or via a webcam.  In most cases, the position and rotation of the eye (or eyes for binocular systems) is determined, and then using additional information on orientation of the user's head, eye tracking data can be converted into gaze tracking.

For the time being, high-quality eye-tracking (accuracy below 1 degree) comes at a financial cost.  However, through the use of machine learning and artificial intelligence techniques, webcam based tracking (currently accurate to around 2-5 degrees) is improving \citep{Harezlak2014}.   As with speech-to-text conversion, low-cost and easy to deploy systems would increase both the scope of user studies in measuring visual discovery skills of astronomers, along with becoming a practical new input device that could enhance visual discovery workflows.   In the next Section, we discuss several user applications in astronomy that could arise by utilising eye and gaze tracking more regularly.

Not only does eye tracking provide information on how astronomers perform visual discovery tasks, it can also be used to determine an individual's cognitive state.   Regardless of skill level or expertise, cognitive factors (tiredness, external distractions, emotional state and mood etc.) can all impact on how, and how effectively, tasks are performed.

\subsection{Cognitive and physical factors}
The longer we complete a repetitive task, the more the cognitive and physical (or psychophysiological) factors are likely to play a part.  Continuous visual inspection and decision-making requires a high-level of mental effort, but is often associated with minimal physical effort.  

Cognitive state, such as attention, can be measured using eye-trackers.  
As \citet{Henderson13} have shown, the task that a viewer is completing while viewing a scene can be determined by eye movements processed using multivariate pattern analyses (e.g.,  scene search, memorisation). Cognitive state is inferred by extension of understanding the mental task being undertaken. In addition, pupillometry has long been used as an indicator of cognitive engagement and effort during task completion \citep{Beatty66}. Specifically, pupil changes are associated with the demands of a task \citep[e.g.][]{Laeng12}.

Additional measurements of physical state [e.g. heart rate and heart rate variability -- see \citet{Laborde17} for a discussion of experimental planning, measurement and analysis] can be obtained with biometric sensors, such as one-lead electrocardiography, or implemented as wearable devices [see the comparative analysis by \citet{Reali19} -- care must be taken if employing or interpreting results from some wearables].  As an example of a study that could be adapted to astronomy, \citet{Laborde15} used heart rate variability to assess the impact of different coping strategies (e.g. emotional intelligence, attention strategy, perceived stress intensity) employed by a cohort of 96 sports science students while performing visual search tasks under pressure.

In a continuous data-streaming/decision-making workflow, tiredness and fatigue will likely arise naturally, and will require careful management.   For shorter  experimental user studies, cognitive fatigue can be induced  through protocols such as the Stroop color word interference test \citep{Stroop35} or the Time load Dual-back paradigm \citep[TLoaDBack; e.g.][]{Borragan16}. Self-reporting of cognitive state can also be used, for example through questionnaires such as the NASA Task Load Index \citep{Hart88}, however, cognitive fatigue is a subjective phenomenon.
 
\section{Applications, opportunities and challenges}
We now look at four potential early-stage applications that could utilise the human performance measures described in Section \ref{sct:cyber}: (1) understanding and improving visual search strategies; (2) talent identification; (3) just-in-time coaching; and (4) adaptive user interfaces.   

We limit our proposed applications to visual-processing tasks, i.e. any action requiring an astronomer to look at a representation of data and identify a feature of interest (i.e. {\em visual discovery}).   Our goal is to build understanding of the nature of expertise and skill in visual discovery amongst astronomers, and assess the impact of cognitive and physical factors, which could impede human performance when working with continuous streaming of data and/or real-time decision-making.

\subsection{Application: Understanding and improving visual search strategies}
\label{sct:sonify}
As astronomy heads closer to the Square Kilometer  Array-era of continuous data-streaming, the efficiency at which astronomers are able to perform visual discovery tasks will rely on both latent skill and expertise.  The paucity of user studies (Section \ref{sct:background}) presents an immediate opportunity to start gathering data on novice and expert performance.  

Utilising the measurements of human performance data introduced in the previous Section, especially eye-tracking and speech-to-text conversion, would allow us to answer questions about visual discovery by novices and experts.  These include: Where do astronomers look when they are examining images? Do they concentrate on the main object? Do they ever look at the whole image, using either a systematic approach or a more haphazard one? How do factors such as  ``dynamic range'' (difference between maximum pixel value and minimum pixel value), noise level, and even colour map change an individual's scanning patterns?  How do eye movements change with increasing skill or expertise, and what is the impact of fatigue?

Access to this knowledge about individuals, or specific cohorts (optical astronomers vs. radio astronomers, early candidature vs. late candidature graduate students, etc.), may lead to new ways to present images that are fine-tuned to maximise discovery potential.

An approach that we intend to investigate in future work is the creation of a set of baseline visual discovery tasks, which could be presented to an international cohort of astronomers through a web application.   Such a set of standardised tasks would enable studies of routine expertise (Section \ref{sct:exskill}). Adaptive expertise, however, requires novel tasks, which might be drawn from outside of the domain of expertise \citep{Carbonell14}, and so is beyond the scope of this current work.

While our user experiments emphasise visual discovery activities, cyber-human discovery systems are likely to be enhanced through multi-sensory input and output.   Data sonification -- mapping features within a dataset to volume, pitch, tonal quality --  has been investigated in astronomy and space physics by, for example, \citet{Candey05,DiazMerced08,DiazMerced12,Tutchton12} and \citet{Cooke19}.  

Sonification often takes adavantage of the auditory system's ability to focus on and isolate low signal-to-noise features, particularly in temporal data streams.   An additional advantage of sonification is that it can provide a more accessible approach to data exploration and analysis for non-sighted researchers.  Identification of, and support for, expertise at multi-sensory investigation is a much broader topic for future work.

\subsection{Application: Just-in-time coaching}
\label{sct:coaching}
How do we coach novices or newcomers to visual discovery in astronomy?    With limited time for training,  does the ``master-apprentice model'' still work, or can we also make use of autonomous systems to identify behaviour and provide just-in-time coaching and training more effectively and efficiently?

With less time spent by humans looking at data, the way astronomers learn to make discoveries or data-driven decisions will need to evolve as well.
Instead of gathering expertise by spending an extended period of graduate study learning the fine details of data reduction with IRAF \citep[Image Reduction and Analysis Facility;][]{Tody86,Tody93},\footnote{https://ascl.net/9911.002}, MIRIAD \citep{Sault95},\footnote{https://www.atnf.csiro.au/computing/software/miriad/} or CASA \citep[Common Astronomy Software Applications][]{McMullin07},\footnote{https://casa.nrao.edu} astronomers of the future will need different skills to ensure that the discoveries, anomalies and artifacts identified by intelligent systems are classified and followed-up correctly \citep{Norris10}.

While astronomers might not be explicitly taught how to view images most effectively, there are additional cultural factors that play a role.  Artists discovered, and are taught, the principles of using composition of features within an image to lead the eye and hold the viewer's attention \citep{English17} -- a strategy used by the Hubble Heritage Team \citep{Rector07}.  See also \citet{Borkin13}, who investigated the related question: ``what makes a visualisation memorable?''

Indeed, it is an open question whether training should be explicit, or could implicit methods be more appropriate and effective? An example here is in terms of providing an increased volume of training, with the expectation that performance may be less likely to ``break down'' under pressure and hence be more robust \citep[e.g.][]{Masters92}.   A related issue is the quantity and nature of feedback that should be provided during training: intuitive decision-making is strengthened through feedback on correctness, but explanations are not essential \citep[e.g.][]{Hogarth08,Schweizer11}.  For performance measures based on eye-tracking, it may be sufficient to provide astronomers with copies of the attention maps and gaze plots generated, such that they can self-assess whether they are looking at the right features in an image.  This approach is in line with the guidance hypothesis \citep{Schmidt91} to avoid overreliance on feedback, and prompt effective learning and problem solving for performance enhancement.
 
By gathering human performance data, it may be possible to identify particular strengths and weaknesses in an individual's approach to visual discovery tasks in real-time.  With this information, it becomes possible to provide tailored coaching, rather than referring the person to undertake self-directed training or further education.

The career trajectories in Figure \ref{fig:ExpertCareer} are not exhaustive, but indicate plausible changes in skill level as career stage progresses, with and without the intervention of user-specific coaching:
\begin{itemize}
\item $A \rightarrow B$: a routine novice, who undergoes no coaching, experiencing a natural decline in skill level over time.   This decline may be due to competing priorities that occur as a career progresses (e.g. less time spent on visual discovery tasks, non-adoption of methodological changes) or  physical factors, such as age-related vision deterioration.
\item $A \rightarrow C$: a routine novice, who is able to unlock increasingly elite potential through self-directed methods, such as seeking alternative approaches to visual discovery.
\item $A \rightarrow D \rightarrow E$: a routine novice who is provided with specific coaching that is able to unlock latent elite potential; 
\item $F \rightarrow G$: a novice with inherent elite skill who receives no additional coaching.
\end{itemize}

In Figure \ref{fig:ExpertCareer}, the Career Stage axis may span a very different range of time for different astronomers.  Indeed, with appropriate coaching, it is likely that expertise could be gained more quickly rather than simply through the passage of time.  Consider the case of medical diagnosis: as hospitals continue to scale-up in terms of the number of patients assessed and treated, there has been a dramatic rise in the rate at which imaging data is presented for review.  Due to the near-continuous nature of new scans and images for reporting, particularly in emergency medicine settings, medical interns (i.e. novices) need to develop expertise much faster in order to ensure accuracy and timeliness in their decisions.\footnote{Dr Peter Santos (Western Health), private communication}

By gathering data on different work patterns, and developing mechanisms for identifying or classifying those into different categories of expertise  (e.g. through machine learning or other forms of artificial intelligence), individuals could be presented with just-in-time training to support them through a particular visual discovery process.   This could be in the form of automated prompts showing how an expert achieved a particular task, such as confirming that the quality of an image from a data reduction work flow was appropriate, or demonstrating how a decision was made that a potential transient source signature belonged to a particular class.

Moreover, the impact of specific coaching strategies could be measured and assessed through a longitudinal study of a cohort of novices.  Here, development in visual discovery skills would be tracked over time via an on-going record of eye movements, verbal reports, etc.

\subsection{Application: Talent identification}
\label{sub:experts}
The ability to provide more targeted or nuanced training to an individual with latent elite skill level relies on identifying that talent.   For competitive sports, talent identification is a skill in its own right, as it often leads directly to a team achieving a winning advantage.  

Currently, there is no understanding as to what characterises an elite talent at visual discovery in astronomy. Is it the ability to work at a higher data rate? Is sustained performance more important than bursts of potential?  Is elite skill linked to coping strategies, emotional regulation, or an ability to sustain attention under pressure?

Through human performance data, we can explore whether there are
indeed identifiable differences in the way that novice and expert astronomers make data-driven, visual discoveries.   This knowledge might then allow for potential elite talent to be identified and nurtured, perhaps leading to the design of better methods to coach and train novices.

\subsection{Application: Adaptive user interfaces}
\label{sct:adaptive}
 When astronomers look at data, they need to see it in the right way and at the right time.   However, not all astronomers will necessarily gain the same insight or make the same decision regarding the nature of a signal, detection or discovery presented by an autonomous workflow. 

A path with future potential is the development of adaptive user interfaces, which ensure that astronomers are provided with the right visualisations automatically.  Such an approach would depend on autonomous systems  that sense an astronomer's mood, cognitive state, or latent skill level as they undertake a visual processing task.   Eye movements, respiration, heart rate and heart-rate variability are all indicators of attention and cognitive state, which can be measured continuously by sensors.  

An adaptive interface might automatically reduce the quantity or nature of information presented as an astronomer becomes fatigued.  Conversely, during elevated periods of attention and cognition, bringing in new or more data might allow for an enhanced discovery potential -- the astronomer is ``in the zone'', or experiencing flow \citep{Nakamura09}, and is more able to link together ideas and generate new knowledge. In this state, performance feels effortless, with high concentration, and a strong feeling of control over the task. An adaptive interface can play a significant role in providing information at a moderate, optimal level of challenge to promote flow and break through, as advocated in learning of other complex tasks \citep[i.e., the ‘challenge point’,][]{Guadagnoli04}.

\begin{table*}
\caption{Six scenarios for assessing user interactions with, and suitability of, data analysis and visualisation environments from \citet{Lam12}.  The seventh scenario, evaluating visualisation algorithms (VA), is less relevant for studying human performance. }
\label{tbl:seven}
    \centering
    \begin{tabular}{c|cc}
    {\bf Code} & {\bf Description} & {\bf Methods} \\
    \hline
        UWP & Understanding environments and work practices & Field observations,
        interviews, \\
        & {\em What are the work or analysis practices that are used?}
        & laboratory observation\\ \hline
        VDAR & Evaluating visual data analysis and reasoning &  Case studies, laboratory observation, \\
        & {\em Does a visualisation tool support development of insight?}
        & interviews, controlled experiments\\ \hline
        CTV & Evaluating communication through visualisation & Controlled experiments, \\
        &  {\em With regards to learning, how effective is a visualisation method? } & field observations and interviews \\ \hline
        CDA & Evaluating collaborative data analysis & Heuristic evaluation, log analysis, \\
        & {\em How effective is the visualisation strategy} & field or laboratory observation, \\
        & {\em at encouraging and supporting collaboration?}  & \\ \hline
        UP & Evaluating user performance &  Controlled experiments,  \\
        & {\em What do objective measurements of user performance (time,  }& field logs \\
        & {\em accuracy) tell us about suitability of a visualisation strategy?} & \\ \hline
        UE & Evaluating user experience & Informal evaluation, usability test, \\
    & {\em What does subjective feedback, written or oral, } & field observation,  \\ 
    & {\em inform us on perceptions of suitabaility/efficiency?} & laboratory questionnaires\\ 
    \hline
    \end{tabular}
\end{table*}

\subsection{Challenges of participatory design and user studies}
\label{sct:seven}

\citet{Lam12}, and details therein, provides a comprehensive overview of the seven main strategies available for assessing user interactions with, and suitability of, data analysis and visualisation environments (summarised in Table \ref{tbl:seven}).  Application of six of the seven strategies is necessary in the development and evaluation of cyber-human discovery systems, sharing many features with co-design or participatory design practices. The seventh scenario, evaluating visualisation algorithms (VA), is less relevant for studying human performance.  Instead, VA examines suitability or quality of a visualisation algorithm, considering factors such as computational efficiency of the implementation.

Both quantitative (e.g. log analysis in CDA and UP scenarios) and qualitative methods (e.g. interviews, questionnaires in UWP, CTV and UE) provide insight on questions of suitability, usability, and effectiveness of visualisation strategies.   Observing how astronomers actually use the solutions they currently have (UWP, UE) in order to gain insight and perform data-driven decision-making needs to occur in tandem with controlled experiments or laboratory observations.

Controlled experiments may, by necessity, result in somewhat contrived investigations. \citet{Meade14} used an artificial task of locating words and astronomical objects in images on both TDWs and desktop displays in a user study containing many more non-astronomers than domain experts.  In comparison, the efficacy of a display ecology that incorporated a TDW was assessed by \citet{Meade17}  during actual DWF observing campaigns.  Thus, a relevant consideration when planning user studies is one of ecological validity:  how real does the experience need to be to have value?

In some disciplines where the use of eye-tracking is well-established, especially sports performance analysis \citep[e.g.][reviews 40 year of eye-tracking research]{Kredel17}, challenges are imposed by the need to track natural gaze behaviour in both ``real world'' and laboratory conditions.  It can be challenging to produce valid laboratory conditions that accurately replicate the visual enviornment and decision-making processes that occur during competitive sport.   For astronomy, where much of the visual discovery work occurs at the desktop, real world and laboratory condition scenarios have substantial overlap.

Perhaps the greatest barrier to user studies is the availability, or willingness, for relevant users -- astronomers -- to participate.   Here, the use of in situ (i.e. field observations) may increase the pool of participants, but with a greater need to integrate additional data gathering solutions (e.g. cameras, microphones, sensors) into a workplace.   One way this can be achieved is by building in human performance monitoring, as has been trialled successfully during Deeper Wider Faster observing campaigns (Hegarty et al. {\em in prep}).

Laboratory observations or controlled experiments require a commitment from astronomers to step aside from their other responsibilities, albeit for a brief period of time, and engage in a user study.  This often occurs at a venue away from their usual place of work, particularly if specific hardware is required as part of the user evaluation.  A compromise may be to provide an online experience, subject to the availability of readily available hardware such as web cameras, computers with in-built microphones, and so on.    Here, astronomers would participate in a set of test scenarios from their office while their interactions with an interface were logged, gaze and attention were tracked, and spoken explanations were recorded.  

For such an approach to be acceptable, and successful, participants would need to have any concerns about privacy allayed -- handing over control over a personal web-camera stream to a third party  service could rightly cause concerns about how that data was being used.

\section{Conclusions}
The role of the astronomer in making discoveries, drawing insight, and generating new knowledge continues to undergo transformations:  from artist to photographer; from working with small digitally-derived data sets to Virtual Observatories connecting geographically distributed archives.   The next transformation is underway, as data volumes and, perhaps more importantly, data velocities exceed the capabilities of individuals or even teams of humans to contend with.   

These changes were driven by the emergence of new technologies: photography (daguerreotypes, wet and dry collodion processes, glass photographic plates, and charge-coupled devices); computers and computing networks; and most recently, the rise of artificial intelligence and machine learning.   Astronomers have adapted in order to capitalise on the opportunities these new technologies bring, in order to advance scientific knowledge of the Universe.

Taking a simplified view of the progress of scientific revolutions \citep{Kuhn1970}, such transformations and transitions do not occur instantaneously, but can extend over time before they are recognised.   Photographic plates and charge-coupled devices co-existed for some time before the former solution entered obsolescence. Additionally, successful elements of former stages can continue to grow in value: management of petabyte and exabyte-scale data collections within well-structured data archives, as per Virtual Observatories, will be an essential component of the infrastructure for cyber-human discovery systems.

As astronomy heads closer to the Square Kilometer Array-era of continuous data-streaming, with an ever more significant reliance on automated processing, the role of the astronomer will evolve.  Effective cyber-human discovery systems will be required, which adapt to the needs, skills, and cognitive state of the individual, while supporting a tighter (human-in-the-loop) working relationship with autonomous systems.

Before proposing any changes in the way astronomers participate in visual discovery or real-time decision making, we need to improve our understanding of current processes. Moreover, it is important to recognise that a variety of skill levels exists amongst the astronomical community: some astronomers are (or could be trained to be) more effective at visual processing tasks than others.  

In this paper, we considered two classes of human factors that may impact on the effectiveness of cyber-human discovery systems in astronomy:  (1) the differences between expertise and skill; and (2) the influence of cognitive and physical factors.  

Automating assessment of skill level, attention, and cognitive state in order to provide  adaptive interfaces (Section \ref{sct:adaptive}) or just-in-time coaching support (Section \ref{sct:coaching}), requires measurement of user workflows and biometric factors.   The former can be achieved by capturing user inputs or interactions with a visualisation tool, which can be easier to achieve for newly-developed applications.   The latter requires additional hardware to perform tasks such as eye-tracking, audio (i.e. speech) recording, or monitoring heart rates, skin conductivity, etc. 

Integration of sensor hardware must be approached in a way that minimises the invasiveness of such measurements, while also establishing (more likely in a controlled experiment) that such additional measurements do provide necessary insight for assessing skill level, attention or cognitive state.   This is the essence of the cyber-human discovery system, where there is a more complete integration between human activities and automation/computation.

Engaging and leveraging the human visual system will remain a fundamental feature of astronomy -- at least for the foreseeable future -- but with digital technologies playing an assistive role.
Human factors such as expertise, skill, cognitive and physical factors all impact on an individual astronomer's ability to work most effectively, efficiently and successfully when aided by automated processes.  Astronomers of the future will not work the same way as astronomers of the past, nor should they.   

\section*{Acknowledgements}
This research has made use of NASA's Astrophysics Data System Bibliographic Services.
The authors thank Shane Vincent and Lewis de Zoete Grundy for their preliminary studies into speech-to-text and web camera eye-tracking. CJF thanks Dr Peter Santos (Western Health) for insightful discussions about expertise and skill in radiological imaging.  The authors thank the two anonymous reviewers.



\bibliography{biblio} 




\end{document}